\documentclass[aps,pre,a4paper,10pt,twoside,twocolumn,showpacs]{revtex4-1}

\usepackage{amssymb,times,graphicx}

\usepackage{hyperref}
\hypersetup{pdftitle={Numerical evidence against a conjecture on the cover time of planar graphs}, pdfauthor={J. Ricardo G. Mendonça}, pdfsubject={Graph Theory, Monte Carlo simulations}, pdfkeywords={Planar graph, cover time, lower bound, random walk, Monte Carlo simulation, probability distribution function}, unicode=true, extension=pdf, pdfnewwindow=true, pdftoolbar=true, pdfmenubar=true, pdffitwindow=false, pdfstartview={FitH}, colorlinks=true, linkcolor=blue, citecolor=blue, urlcolor=blue}

\def\leq{\leqslant}
\def\geq{\geqslant}

\begin{document}

\title{Numerical evidence against a conjecture on the cover time of planar graphs}

\author{J. Ricardo G. Mendon\c{c}a}
\email[Email: ]{jricardo@if.usp.br}
\affiliation{Instituto de F\'{\i}sica, Universidade de S\~{a}o Paulo -- Caixa Postal 66318, CEP 05314-970 S\~{a}o Paulo, SP, Brazil}

\begin{abstract}
We investigate a conjecture on the cover times of planar graphs by means of large Monte Carlo simulations. The conjecture states that the cover time $\tau(G_{N})$ of a planar graph $G_{N}$ of $N$ vertices and maximal degree $d$ is lower bounded by $\tau(G_{N}) \geq C_{d}N(\ln N)^2$ with $C_d = (d/4\pi) \tan (\pi/d)$, with equality holding for some geometries. We tested this conjecture on the regular honeycomb ($d=3$), regular square ($d=4$), regular elongated triangular ($d=5$), and regular triangular ($d=6$) lattices, as well as on the nonregular Union Jack lattice ($d_{\rm min}=4$, $d_{\rm max}=8$). Indeed, the Monte Carlo data suggest that the rigorous lower bound may hold as an equality for most of these lattices, with an interesting issue in the case of the Union Jack lattice. The data for the honeycomb lattice, however, violates the bound with the conjectured constant. The empirical probability distribution function of the cover time for the square lattice is also briefly presented, since very little is known about cover time probability distribution functions in general.
\end{abstract}



\maketitle


\section{Introduction}
\label{intro}

The cover time of a graph is a classic problem in theoretical computer science and graph theory with many practical implications, e.g., in the development of query processing and routing algorithms in computer networks and distributed systems, and has attracted the attention of computer scientists, mathematicians, and physicists for more than 30 years \cite{focs79,aldous83,%
matthews,jtp89,aldous91,zucker,lovasz,chandra,rwgraphs,kahn,nemirov,brummel,%
kaline,palacios,feige,wijland,jonsch,yokoi,dprz,mixing,zlatanov,blanket}.

For a finite, connected graph $G_{N} = (V,E)$ of order $N$, the cover time $\tau(G_{N})$ is the maximum expected time over the possible starting vertices $v \in V$ it takes for a random walker jumping through the edges of $G_{N}$ with uniform probabilities to visit every vertex of $G_N$ at least once. Exact expressions for the cover time are rare except for the simpler graphs, e.g., for the complete graph, for which the problem reduces to the well-known coupon collector's problem, and for the path, cycle, and star graphs, among a few others \cite{palacios,rwgraphs,zlatanov,mixing}.

Most results on graph cover times come in the form of bounds on their expectation values, although little is known about the limit distributions of the related quantities.  The existence of a lower bound follows from $\tau(G_{N}) \geq N$---although it can be proved that $\mathbb{P}(\tau(G_{N}) \leq cN) \leq e^{-{\alpha}N}$, with $\alpha > 0$ depending only on $c$ and $d_{\rm max}(G_{N})$, the maximal degree of the graph \cite{itai}---, while the existence of an upper bound follows from the recurrence of the associated Markov chains.

For planar graphs, Jonasson and Schramm showed that \cite{jonsch}
\begin{equation}
\label{bound}
\liminf_{N \to \infty} \frac{\tau(G_{N})}{N(\ln N)^2} \geq C_d,
\end{equation}
where $C_{d} > 0$ are universal constants depending only on $d \geq d_{\rm max}(G_{N})$. The functional form $\tau(G_{N}) \sim N(\ln N)^2$ for the cover time of the square lattice had been guessed earlier on the basis of Monte Carlo simulations and scaling analysis, where a multiplicative correction $(1+c/\ln N)$ to this form was detected, with $c$ (the magnitude of the leading scaling correction) a constant depending on the boundary conditions of the finite graphs \cite{nemirov,brummel,kaline}. Following a sophisticated probabilistic-geometric analysis of the cover time of the square lattice, Dembo {\it et al.\/} \cite{dprz} conjectured that for the honeycomb, square, and triangular lattices, corresponding respectively to $d = d_{\rm max} = 3$, $4$, and $6$, the constants $C_{d}$ appearing in Eq.~(\ref{bound}) are exactly given by
\begin{equation}
\label{const}
C_{d} = \frac{d}{4\pi} \tan (\frac{\pi}{d}),
\end{equation}
further conjecturing that, with this $C_{d}$, inequality (\ref{bound}) may actually hold as an equality for these geometries.

In this article we investigate numerically the above mentioned conjecture on some planar graphs to check whether the bound (\ref{bound}) with the conjectured constant Eq.~(\ref{const}) holds as a lower bound or as an equality. As we shall see, the empirical data for the cover time of the honeycomb lattice indicate a violation of the lower bound with the constant given by Eq.~(\ref{const}), requiring a smaller constant, while for some other planar graphs the conjecture seems to hold valid.


\section{Monte Carlo data}
\label{mcdata}

\begin{figure*}
\centering
\begin{tabular}{c@{\hspace{2em}}c@{\hspace{2em}}c@{\hspace{2em}}c@{\hspace{2em}}c}
\includegraphics[viewport=123 203 471 632,scale=0.22,angle=-90,clip]{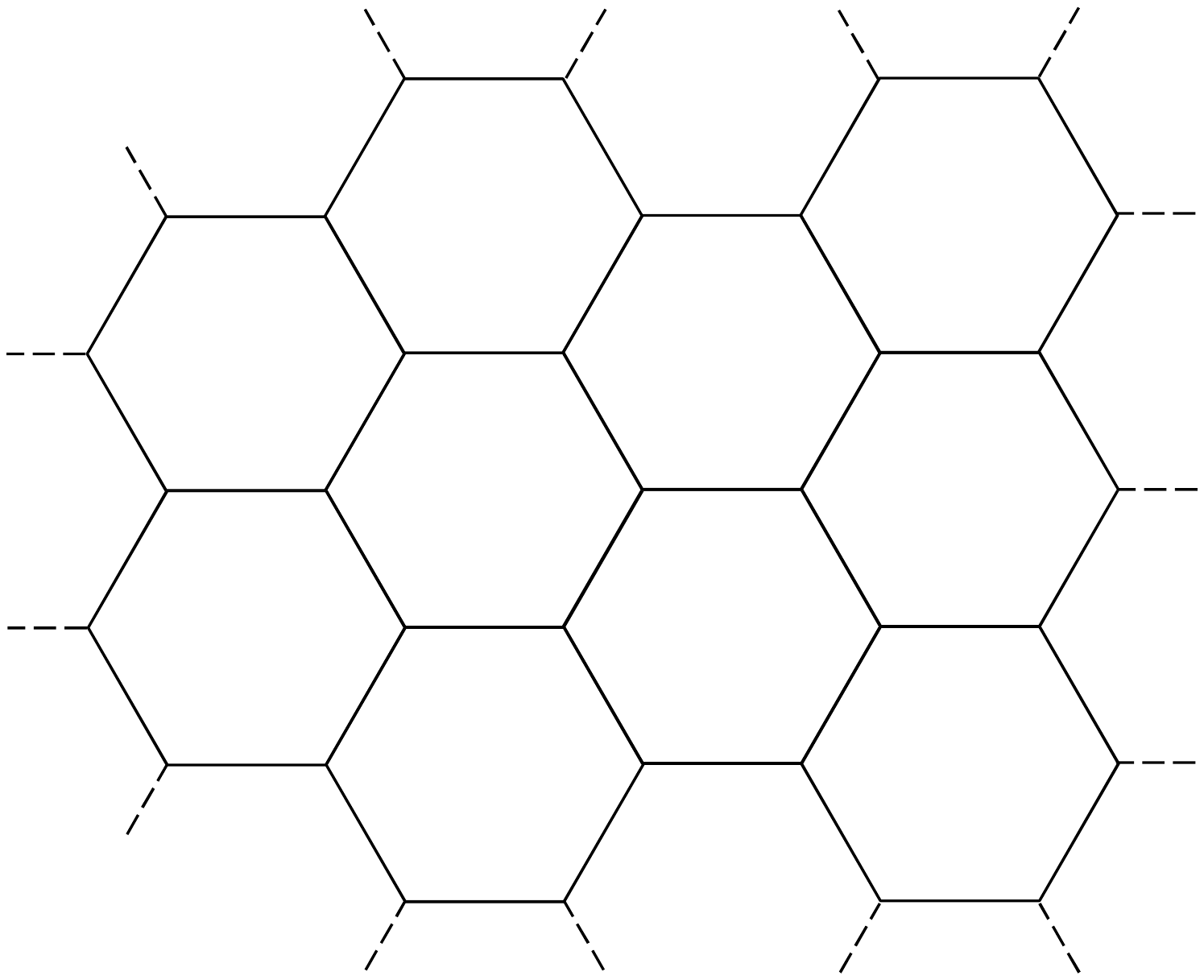}  &
\includegraphics[viewport=155 278 435 564,scale=0.26,angle=-90,clip]{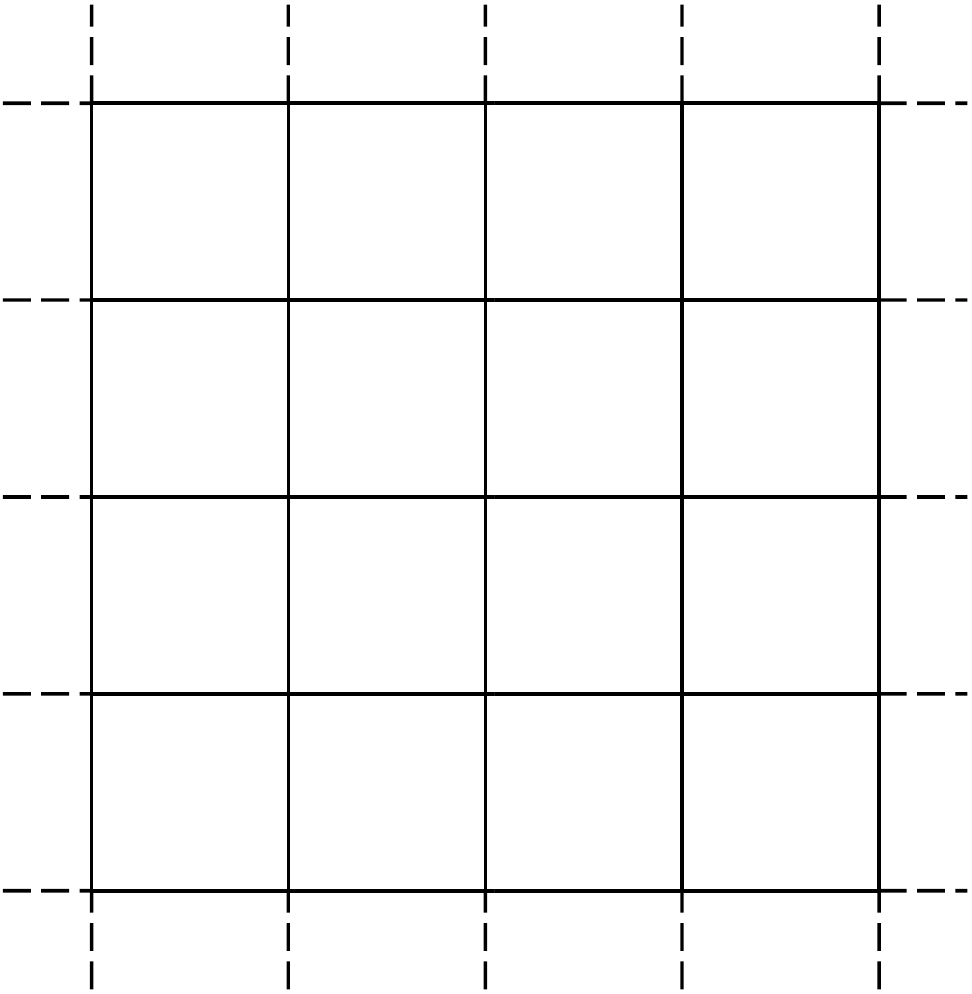} &
\includegraphics[viewport=126 220 442 646,scale=0.26,angle=-90,clip]{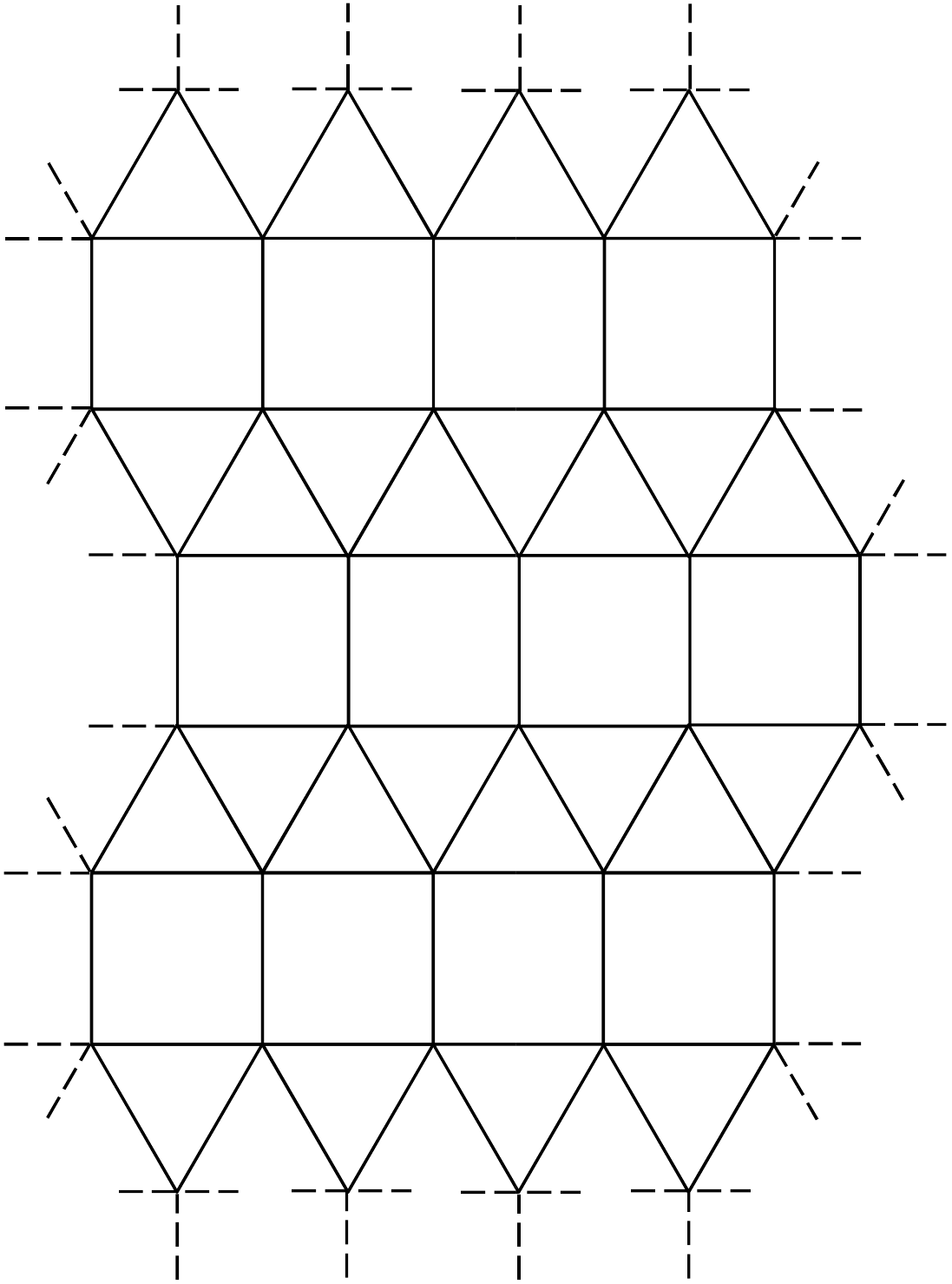} &
\includegraphics[viewport=171 276 423 564,scale=0.28,angle=-90,clip]{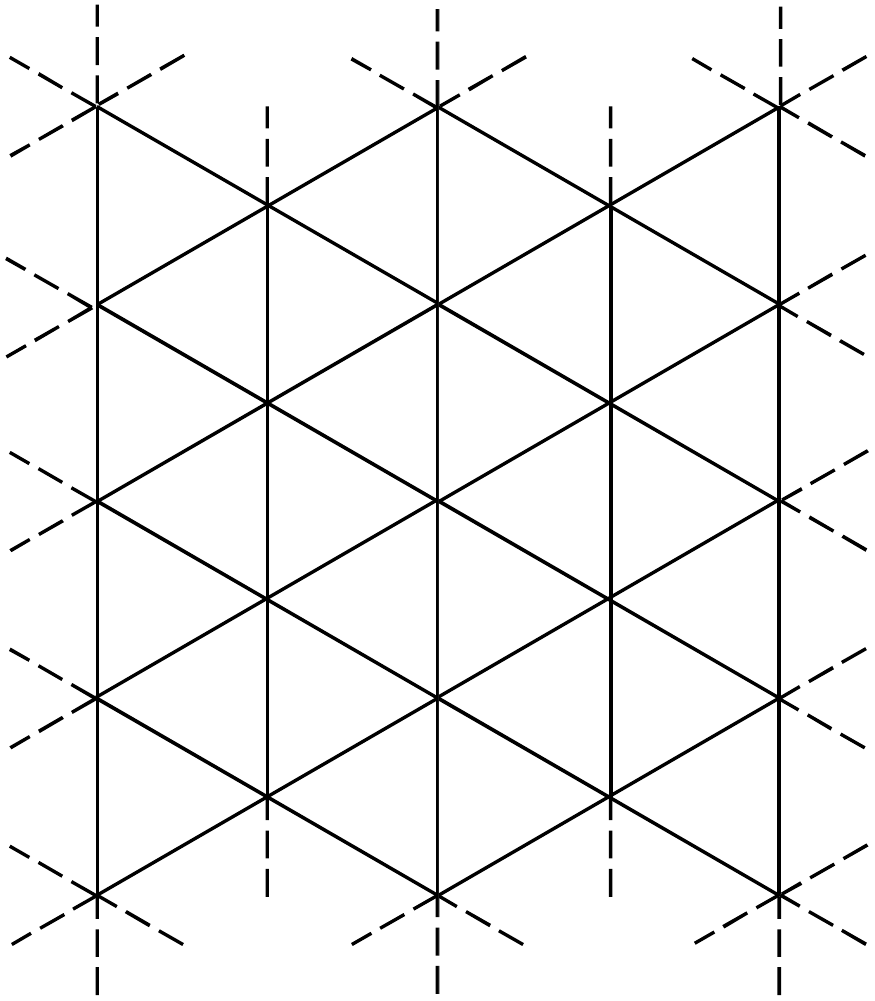} &
\includegraphics[viewport=154 278 441 564,scale=0.26,angle=-90,clip]{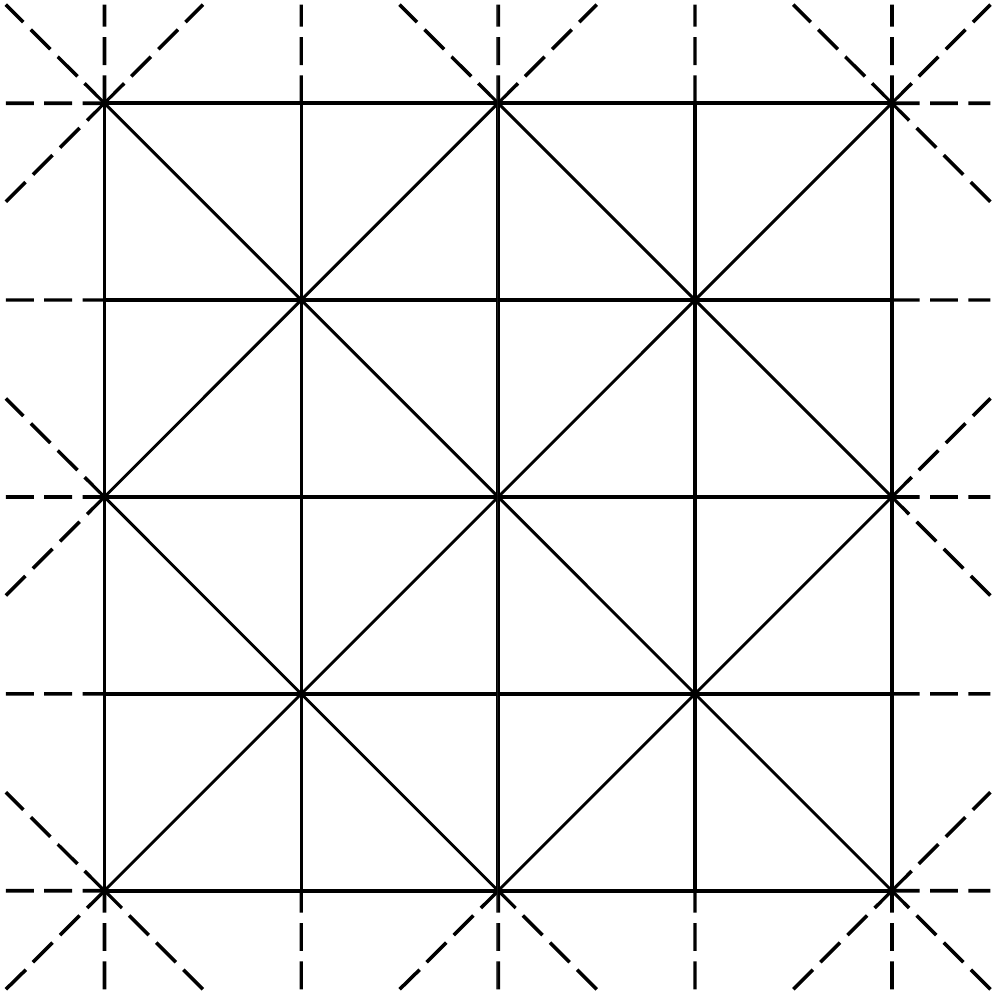}
\end{tabular}
\caption{\label{graphs}Planar graphs investigated in this article. From left to right we have the honeycomb ($d=3$), square ($d=4$), elongated triangular ($d=5$), triangular ($d=6$), and Union Jack ($d_{\rm min}=4$, $d_{\rm max}=8$) lattices.}
\end{figure*}

We investigate the conjecture encoded in Eqs.~(\ref{bound})--(\ref{const}) on the regular honeycomb ($d=3$), regular square ($d=4$), regular elongated triangular ($d=5$), and regular triangular ($d=6$) lattices, and also on the nonregular Union Jack lattice ($d_{\rm min}=4$, $d_{\rm max}=8$). These lattices are depicted in Figure~\ref{graphs}.

\begin{figure}
\centering
\begin{tabular}{c}
\includegraphics[viewport=95 51 447 786,scale=0.285,angle=-90,clip]{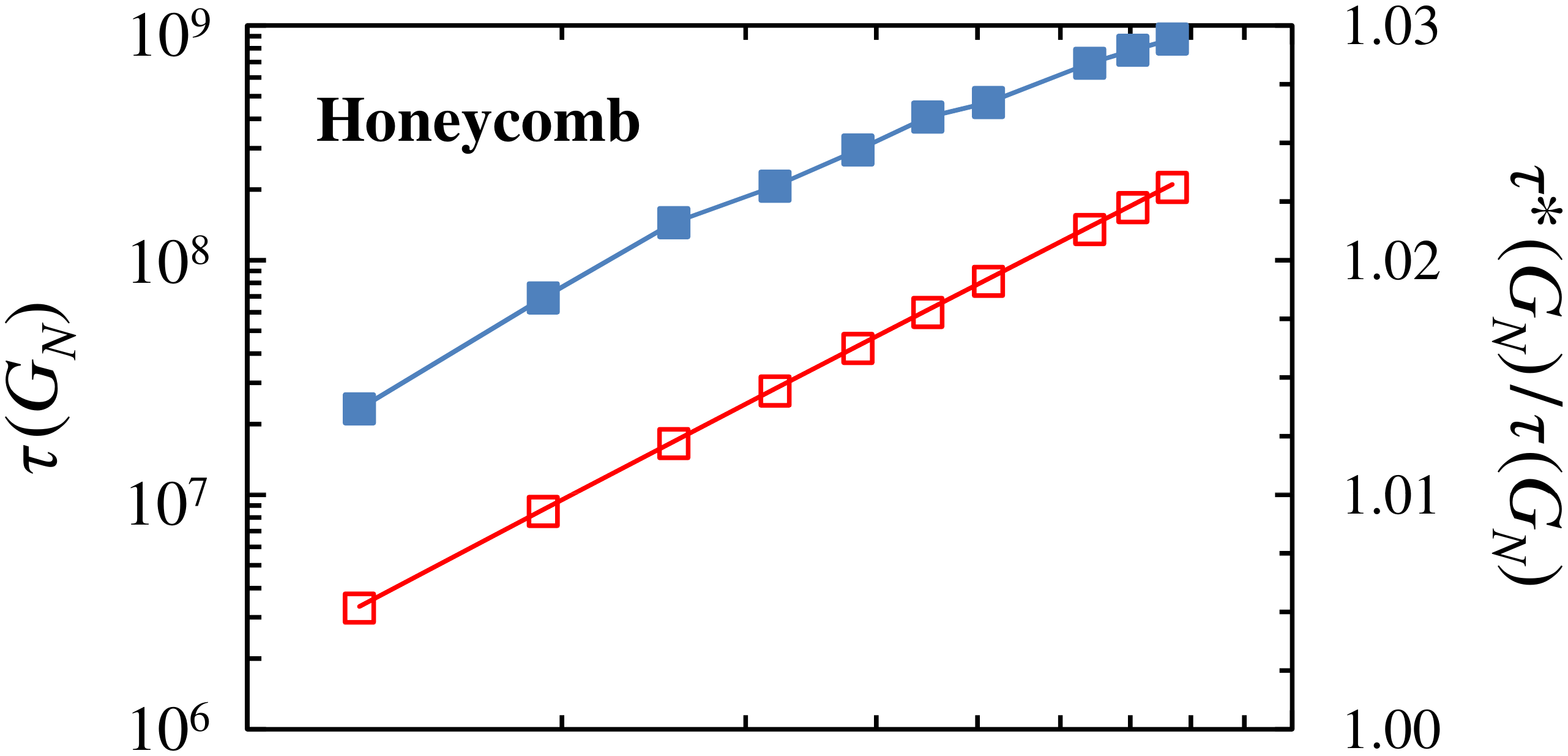} \\
\includegraphics[viewport=95 51 448 786,scale=0.285,angle=-90,clip]{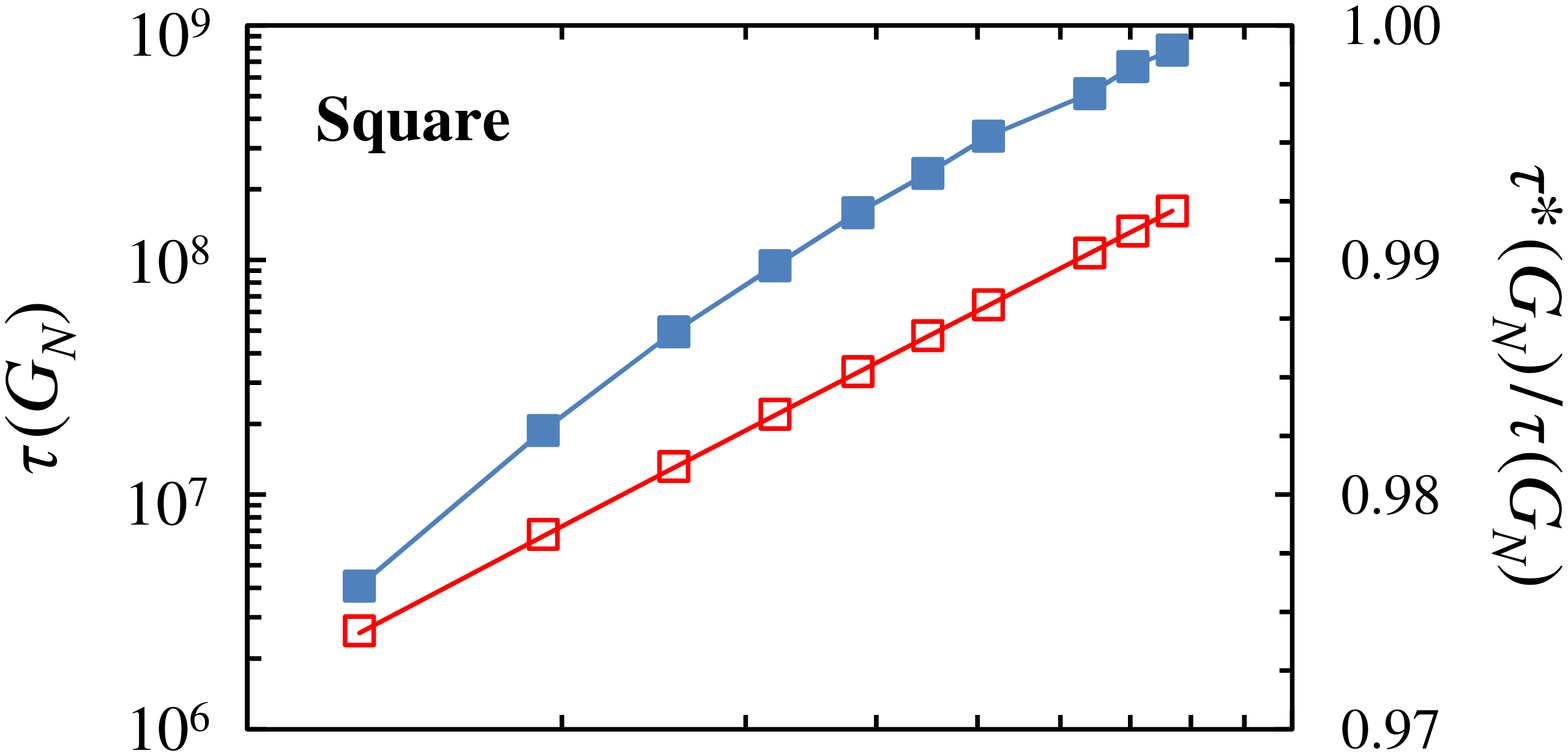} \\
\includegraphics[viewport=96 51 447 786,scale=0.285,angle=-90,clip]{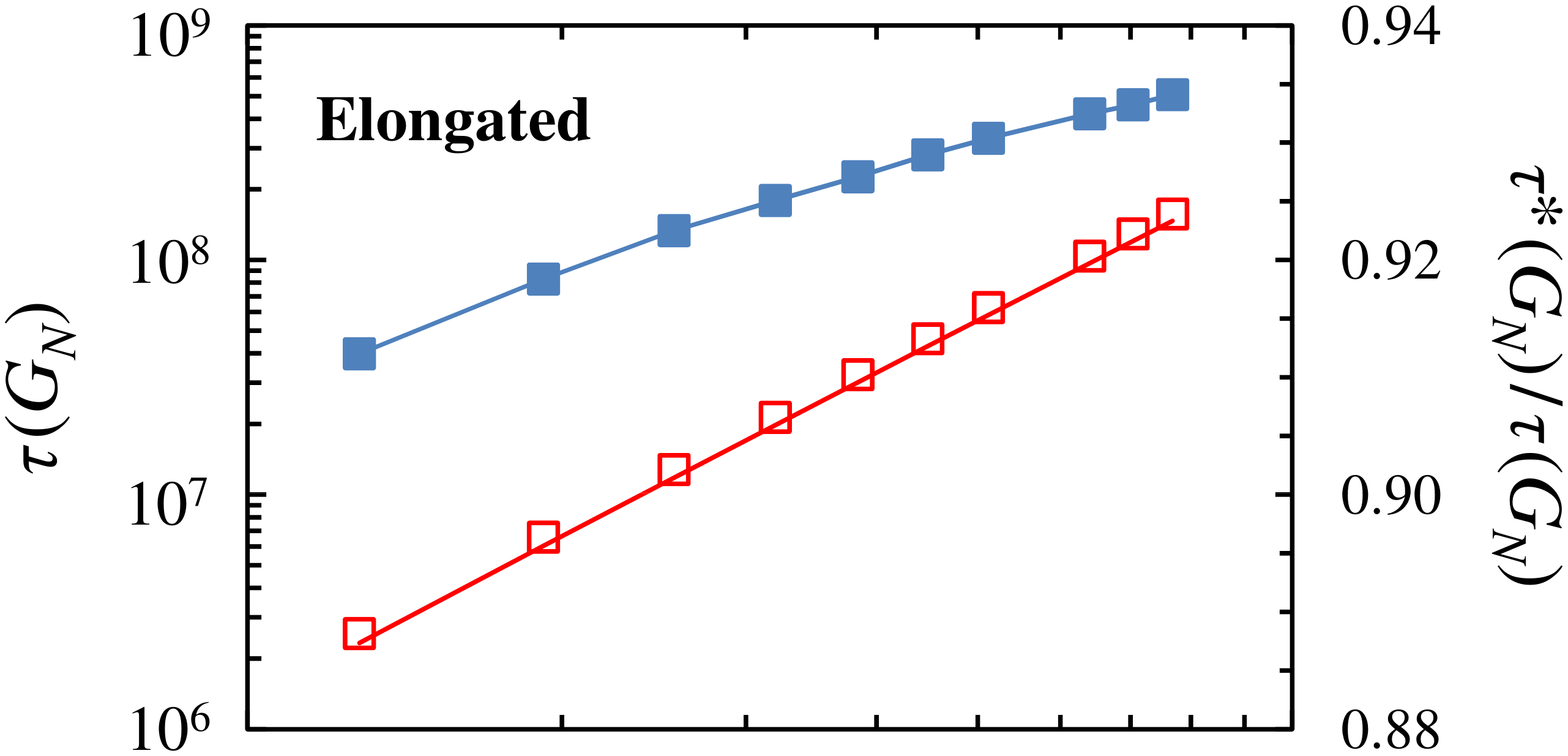} \\
\includegraphics[viewport=96 51 448 786,scale=0.285,angle=-90,clip]{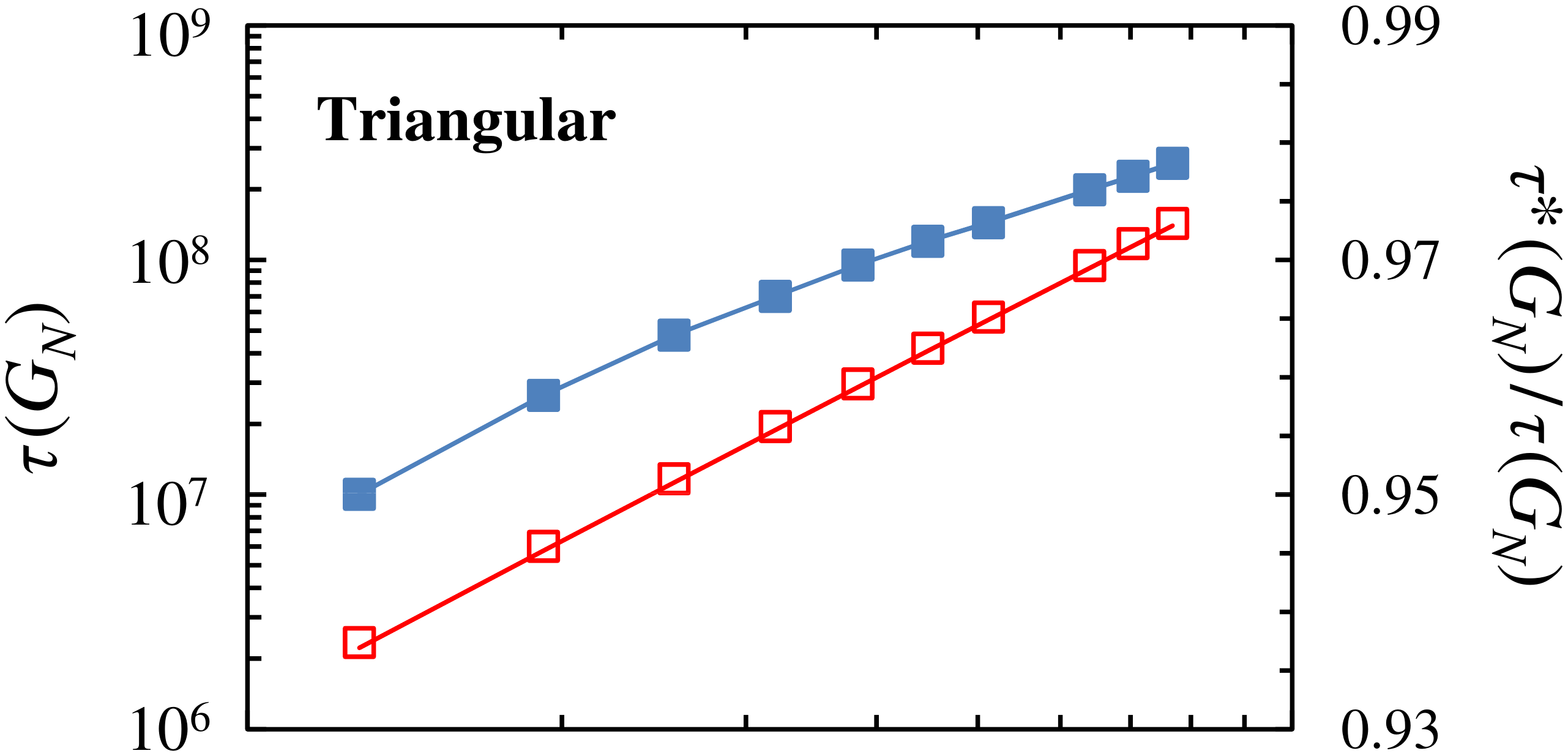} \\
\includegraphics[viewport=96 51 502 786,scale=0.285,angle=-90,clip]{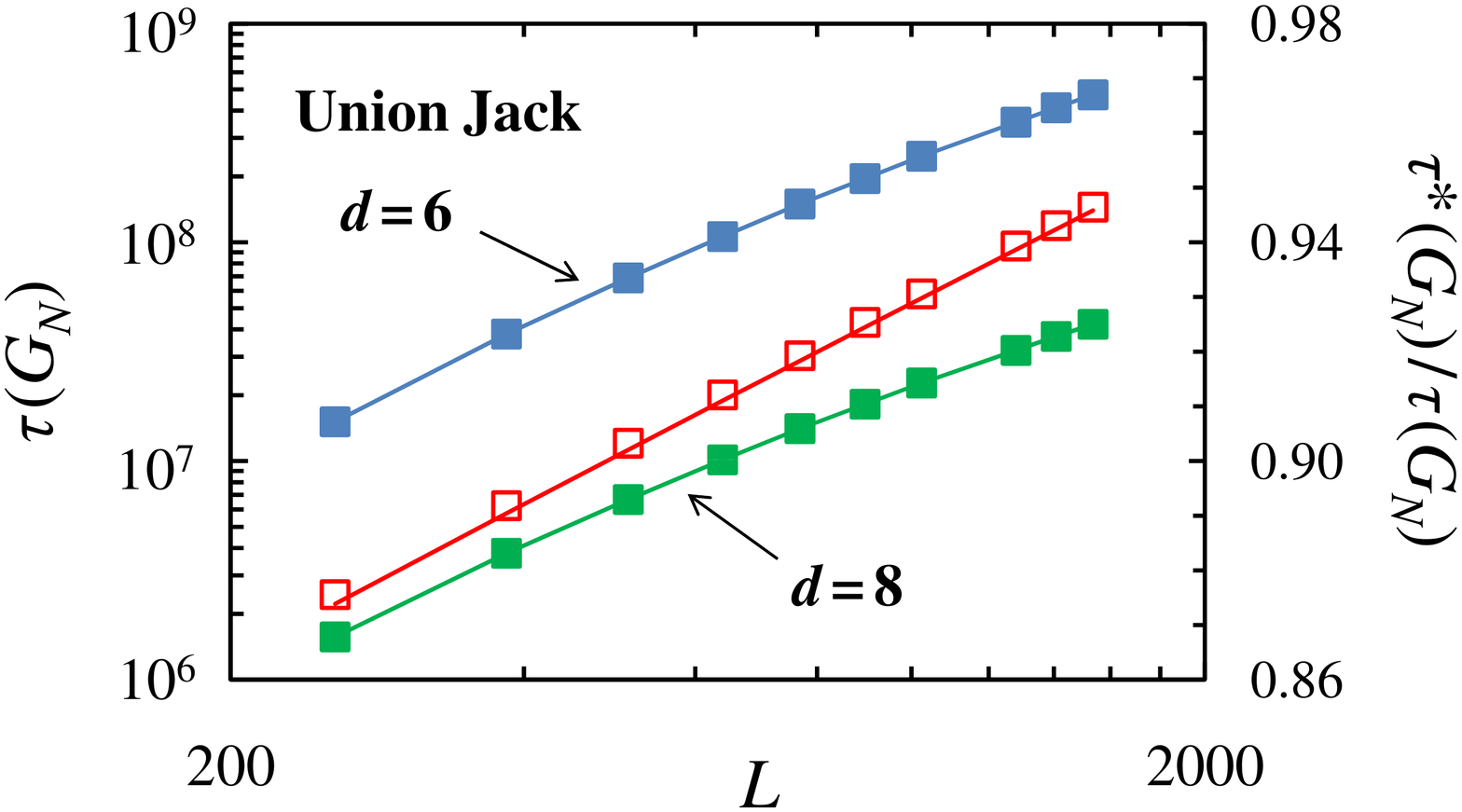} 
\end{tabular}
\caption{\label{taus}(Color online) Cover times for the planar lattices depicted in Fig.~\ref{graphs}. Each graph displays the empirical cover times $\tau(G_{N})$ (open squares, left scale) together with the conjectured value $\tau^{*}(G_{N}) = C_{d}N(\ln N)^{2}$ (solid line, left scale) and the ratio $\tau^{*}(G_{N})/\tau(G_{N})$ (full squares, right scale). Each point of the $\tau(G_{N})$ curves was obtained as an average over $10^6$ samples. For the Union Jack lattice we display the ratio $\tau^{*}(G_{N})/\tau(G_{N})$ both for $C_{d}$ with $d = d_{\rm max} = 8$ and $d = \bar{d} = \frac{1}{2}(d_{\rm min}+d_{\rm max}) = 6$; the plots of $\tau^{*}(G_{N})$ for these two values of $d$ are indistinguishable on the left scale of the graph and appear as a single line passing through the open squares.}
\end{figure}

We computed the cover times on graphs with $N = L \times L$ vertices under periodic boundary conditions, with $256 \leq L \leq 1536$, i.e., on graphs with up to $2.359 \times 10^{6}$ vertices. For each graph geometry and size, $\tau(G_{N})$ is obtained as an average over $10^6$ samples. Our Monte Carlo data together with the conjectured values $\tau^{*}(G_{N}) = C_{d}N(\ln N)^{2}$ obtained from Eqs.~(\ref{bound})--(\ref{const}) appear in Figure~\ref{taus}.

Notice that the toroidal graphs obtained under periodic boundary conditions are not planar, although they are locally very close to planar. The really important fact in a planar graph for the cover time problem, however, is that its edges do not cross, not that it can be embedded in a plane. Moreover, finite graphs with open boundary conditions cannot be regular, since the vertices at the boundaries are of a smaller degree. The asymptotics in the two cases (open and periodic boundary conditions) are expected to be the same, and most results on graph cover times, including those to which we want to compare our own results, are obtained for graphs under periodic boundary conditions.

According to the conjecture, we must observe $\tau^{*}(G_{N})/\tau(G_{N}) \leq 1$ for all planar graphs, a condition that our numerical data does not support for the honeycomb lattice. Other lattices observe the bound, with the square lattice being ``borderline.''  A na\"{\i}ve extrapolation of the ratios $\tau^{*}(G_{N})/\tau(G_{N})$ would give an extrapolated value greater than $1$ in almost all cases depicted in Fig.~\ref{taus}. This indicates that possible corrections to $\tau^{*}(G_{N})$ must go in the direction of decreasing its value. Since previous results in the literature suggest just the opposite, i.e., that, if anything, $\tau^{*}(G_{N})$ may be missing a $(1+c/\ln N)$ correction with $c>0$ \cite{nemirov,brummel,kaline}, we are led to believe that the constant $C_{d}$ with $d = d_{\rm max}$ is overshooting. Supplementary evidence comes from the behaviour of the Union Jack lattice with respect to $d$. While the bound (\ref{bound}) requires $d \geq d_{\rm max}(G_{N})$, our data suggest that this requirement is probably not optimal. We plot the ratios $\tau^{*}(G_{N})/\tau(G_{N})$ for the Union Jack lattice both with $d=d_{\rm max} = 8$ and with $d = \bar{d} = \frac{1}{2} (d_{\rm min}+d_{\rm max}) = 6$, the average degree of the lattice, and we found that the ratio with $d = \bar{d}$ provides a better lower bound than the ratio with $d = d_{\rm max}$; see Fig.~\ref{taus}. This makes us wonder if the average vertex degree
\begin{equation}
\label{avgdeg}
\bar{d}(G) = \frac{1}{|G|}\sum_{v \in V(G)} d(v),
\end{equation}
where $d(v)$ is the degree of vertex $v$, is not a better constant to be used on purportedly universal formulas for planar graphs than the maximum degree $d_{\rm max}$. This could be tested on planar random graphs---e.g., on Delaunay triangulations of random points on the plane \cite{tessel}---, for which $\bar{d}$ can assume different values, integer or not. Recently, planar random graphs have attracted the attention of physicists and mathematicians interested in their connectivity and percolation properties \cite{wierman,bollobas,dickman,hilhorst,neher,ziff}, but their cover times remain unexamined.


\section{The empirical probability distribution function of the cover times of the square lattice}
\label{empirical}

Very little is known about the probability distribution function (PDF) of $\tau(G_{N})$. Actually, it is an open problem to prove that $\tau(G_{N})$ has a nondegenerate limit law \cite{rwgraphs,dprz,mixing}. It seems that the only result on this regard available to date is a concentration result that states that, under mild conditions, the cover time is well approximated by its expected value as $N \to \infty$ \cite{aldous91}. It is thus of some interest to explore our empirical data to characterize the PDF of the cover time, although we will not attempt to identify or infer this distribution here.

\begin{figure}
\centering
\includegraphics[viewport=136 54 459 786,scale=0.33,angle=-90,clip]{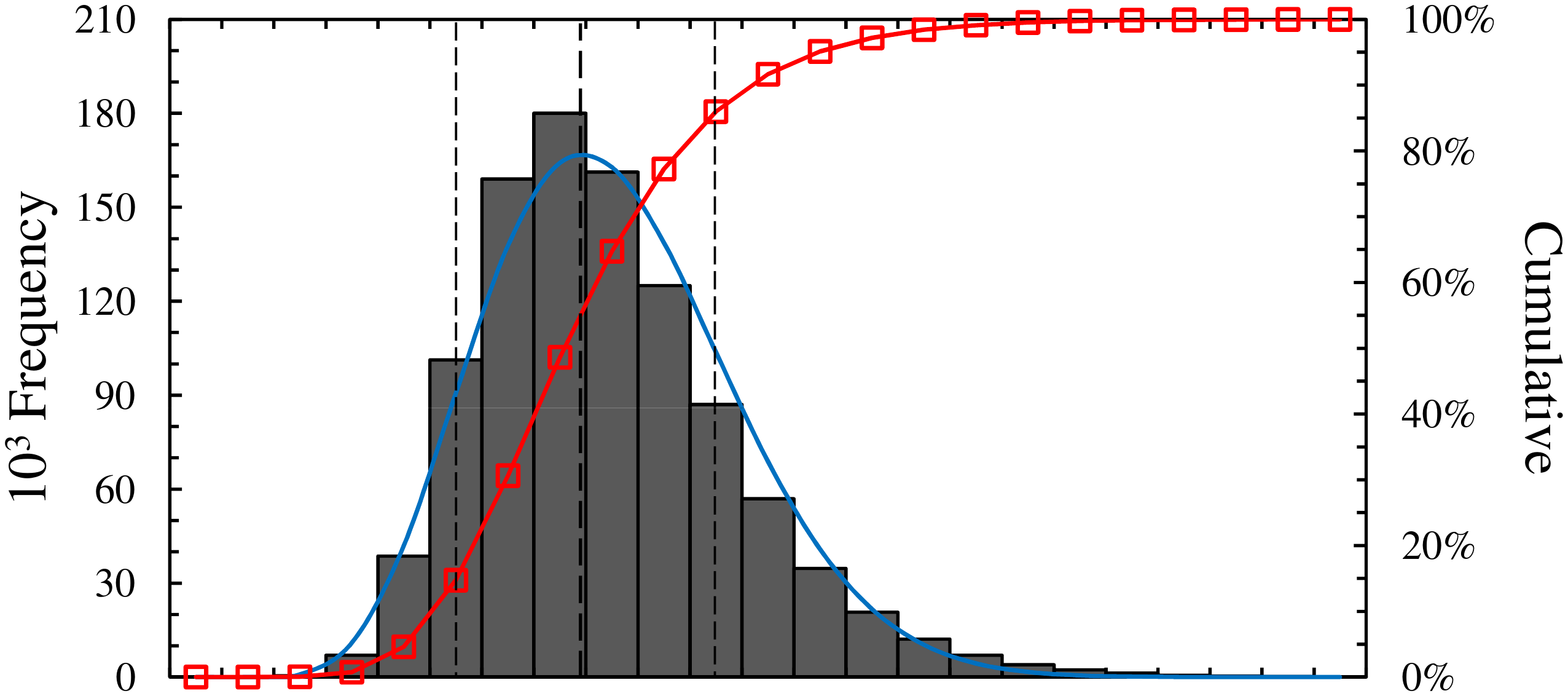} 
\caption{\label{histogram}(Color online) Histogram plot of $10^{6}$ cover times sampled for a square lattice of $ N = 1280 \times 1280$ vertices. There are $23$ bins in the histogram, with the leftmost bin centered at $70 \times 10^{6}$ and the righmost bin centered at $180 \times 10^{6}$. The vertical dashed lines indicate the empirical mean $m$ and $\pm s$ intervals. The continuous line corresponds to the adjusted beta PDF (\ref{beta}) with shape parameters $\alpha = 5.759 \pm 0.008$ and $\beta = 18.04 \pm 0.02$. The empirical cumulative distribution function is also shown (open squares, right scale).}
\end{figure}

Figure~\ref{histogram} shows the histogram plot of $10^{6}$ cover times sampled for a square lattice of $ N = 1280 \times 1280$ vertices. For these data, we compute the sample mean $m = \langle \tau_{i}(G_{N}) \rangle$ and first few central moments $m_{k} = \langle \big(\tau_{i}(G_{N})-m\big)^{k} \rangle$, from which we compute the sample standard deviation $s = \sqrt{m_{2}}$, skewness $g_{1} = m_{3}/m_{2}^{3/2}$, and excess kurtosis $g_{2} = m_{4}/m_{2}^{2}-3$ \cite{stats}. The values of these quantities are collected in Table~\ref{table}.

\begin{table}
\begin{center}
\begin{tabular}{l@{\hspace{2em}}c@{\hspace{2em}}r}
\hline \hline
mean            & $m$     &  $107\,111\,924$ \\
variance        & $s$     &  $12\,308\,707$  \\
skewness        & $g_{1}$ &  $0.91374$       \\
excess kurtosis & $g_{2}$ &  $1.51095$       \\
\hline \hline
\end{tabular}
\caption{\label{table}First moments of the empirical probability distribution function of the cover time of the square lattice depicted in Fig.~\ref{histogram}.}
\end{center}
\end{table}

We also fitted the data to the beta PDF given by
\begin{equation}
\label{beta}
P(x;\alpha,\beta) = 
\frac{\Gamma(\alpha+\beta)}{\Gamma(\alpha)\Gamma(\beta)} x^{\alpha-1}(1-x)^{\beta-1}
\end{equation}
rescaled to the interval $[a,b]$  with $a = \min\{\tau_{i}(G_{N})\} = 73\,391\,036$ and $b = \max\{\tau_{i}(G_{N})\} = 213\,039\,197$. We choose the beta distribution because it has finite support and can take several different shapes; moreover, beta-like PDFs for the cover times of some special graphs were found in previous investigations \cite{nemirov,brummel,zlatanov}. The empirical data together with the adjusted beta PDF appear in Fig.~\ref{histogram} \cite{rmass}.

The positive skewness $g_{1}$ indicates that the empirical PDF is right-tailed, with the bulk of the observed values lying to the left of the mean, although this feature is not very clear from the histogram (\ref{histogram}) because $g_{1}$ is not very large. The moderately high value of the excess kurtosis $g_{2}$, in turn, indicates that the empirical PDF is markedly non-normal, with a sizeable proportion of the data in its right tail contributing to the variance observed. We notice that after $10^{6}$ samplings, the ratio $s/m$ has stationed at $\sim 11.5\%$.


\section{Conclusions}
\label{summary}

Our numerical data for the cover time of the honeycomb lattice provide evidence against the conjecture set forth in \cite{dprz} regarding the constant $C_{d}$ appearing on the lower bound (\ref{bound}) for planar graphs. Otherwise, for the other lattices investigated in this article the functional form given by Eqs.~(\ref{bound})--(\ref{const}) seems to hold valid, possibly as an equality. In summary, for $d=3$ our data seem to falsify the conjecture, for $d=4$ it is ``borderline,'' and for $d \geq 5$ the conjecture holds easily. That there must be something special about the $d=4$ case has been long recognized \cite{rwgraphs}, and this might have showed in our finite-size simulations. Notice that the lower bound (\ref{bound}) remains valid for some $C_{d}$, just not with the $C_{d}$ given by Eq.~(\ref{const}) with $d = d_{\rm max}$. Since the $C_{d}$ given by Eq.~(\ref{const}) is monotone decresing in $d$, it can be used in Eq.~(\ref{bound}) to validate the bound, but then necessarily with some $d > d_{\rm max}$.

It can be argued that our data were taken too short from the limit $N \to \infty$. However, if the conjecture is to be saved, that would mean a nonmonotone convergence of the ratios $\tau^{*}(G_{N})/\tau(G_{N})$, something that our data do not indicate.

Any case for the average degree $\bar{d}$ given by Eq.~(\ref{avgdeg}) would be welcome. It may be that the situation here is similar to that of the determination of the critical percolation threshold $p_{c}$, for which good approximations and scaling relations were found based on the mean Euler characteristic of the critical percolation patterns \cite{neher}. It should be remarked, however, that for the critical percolation threshold problem one does not expect to find a ``universal'' formula in terms of the maximal vertex degree $d$ alone, simply because of the empirical observation that several different lattices with the same $d$ have disparate $p_{c}$. Notice that the constant $C_{d}$ in Eq.~(\ref{const}) is closely related with the filling factor $f$ introduced by Suding and Ziff in order to relate $p_{c}$ to the number of sites per unit area of Archimedean lattices \cite{suding}, 
\begin{equation}
f = \pi \left[ \sum_{i}a_{i} \cot (\frac{\pi}{n_{i}}) \right]^{-1},
\end{equation}
where $(n_{1}^{a_{1}},n_{2}^{a_{2}},\ldots)$ is the Gr\"{u}nbaum-Shephard representation of the Archimedean lattice---e.g., the square lattice is denoted by $(4^{4})$, while the elongated triangular lattice is denoted by $(3^{3},4^{2})$ \cite{tilings}. The investigation of different planar graphs of same degree (regular or average)---e.g., for $d=4$, the square, kagom\'{e}, and Archimedean $(3,4,6,4)$ lattices---may help to elucidate the above questions of geometric character.

Finally, our exploratory analysis of the PDF of the cover times of the square lattice in Sec.~\ref{empirical} is admittedly jejune; that was not the focus of this work. We were nevertheless able to establish that the empirically observed PDF is not normal, with a leptokurtic shape ($g_{2} > 0$) . A proper investigation of the PDF of the cover times of planar graphs---e.g., by model selection among candidate left and right limited, univariate distributions---is still lacking and provides an interesting direction for further statistical work on cover times.


\acknowledgments

The author is indebted to Professor Robert M. Ziff (U.~Michigan) and Professor Henk J. Hilhorst (U.~Paris-Sud) for several constructive comments on a previous version of the manuscript. This work was partially supported by Conselho Nacional de Desenvolvimento Cient\'{\i}fico e Tecnol\'{o}gico -- CNPq, Brazil, through the PDS grant 151999/2010-4.


\end{document}